\documentclass[manuscript]{acmart}
\usepackage{graphicx}
\AtBeginDocument{%
  }

\copyrightyear{2024} 
\acmYear{2024} 
\setcopyright{rightsretained} 
\acmConference[MUM '24]{International Conference on Mobile and Ubiquitous
Multimedia}{December 1--4, 2024}{Stockholm, Sweden}
\acmBooktitle{International Conference on Mobile and Ubiquitous Multimedia
(MUM '24), December 1--4, 2024, Stockholm,
Sweden}\acmDOI{10.1145/3701571.3703381}
\acmISBN{979-8-4007-1283-8/24/12}

\usepackage{subcaption}  

\begin{document}


\title{The Framework of NAVIS: Navigating Virtual Spaces with Immersive Scooters}
\author{Zhixun Lin}
\orcid{0009-0009-2380-2050}
\affiliation{%
  \institution{The Hong Kong University of Science and Technology (Guangzhou)}
  \city{Guangzhou}
  \country{China}
}
\email{zlin719@connect.hkust-gz.edu.cn}

\author{Wei He}
\orcid{0009-0001-6745-2752}
\affiliation{%
  \institution{The Hong Kong University of Science and Technology (Guangzhou)}
  \city{Guangzhou}
  \country{China}
}
\email{danielweihe@hkust-gz.edu.cn}

\author{Xinyi Liu}
\orcid{0009-0006-0788-6619}
\affiliation{%
  \institution{The Hong Kong University of Science and Technology (Guangzhou)}
  \city{Guangzhou}
  \country{China}
}
\email{xliu012@connect.hkust-gz.edu.cn}

\author{Mingchen Ye}
\orcid{0009-0001-2690-1160}
\affiliation{%
  \institution{The Hong Kong University of Science and Technology (Guangzhou)}
  \city{Guangzhou}
  \country{China}}
\email{mye041@connect.hkust-gz.edu.cn}

\author{Xiang Li}
\orcid{0000-0001-5529-071X}
\affiliation{%
  \institution{Leverhulme Centre for the Future of Intelligence, University of Cambridge}
  \city{Cambridge}
  \country{United Kingdom}}
\email{xl529@cam.ac.uk}

\author{Ge Lin Kan}
\orcid{0000-0002-4422-9531}
\affiliation{%
  \institution{The Hong Kong University of Science and Technology (Guangzhou)}
  \city{Guangzhou}
  \country{China}}
\email{gelin@ust.hk}

\renewcommand{\shortauthors}{Zhixun Lin, et al.}

\begin{abstract}
Virtual reality (VR) environments have greatly expanded opportunities for immersive exploration, yet physically navigating these digital spaces remains a significant challenge. In this paper, we present the conceptual framework of \textsc{NAVIS} (Navigating Virtual Spaces with Immersive Scooters), a novel system that utilizes a scooter-based interface to enhance both navigation and interaction within virtual environments. \textsc{NAVIS} combines real-time physical mobility, haptic feedback, and CAVE-like (Cave Automatic Virtual Environment) technology to create a realistic sense of travel and movement, improving both spatial awareness and the overall immersive experience. By offering a more natural and physically engaging method of exploration, \textsc{NAVIS} addresses key limitations found in traditional VR locomotion techniques, such as teleportation or joystick control, which can detract from immersion and realism. This approach highlights the potential of combining physical movement with virtual environments to provide a more intuitive and enjoyable experience for users, opening up new possibilities for applications in gaming, education, and beyond.

\end{abstract}


\begin{teaserfigure}
    \centering
    \includegraphics[width=0.7\linewidth]{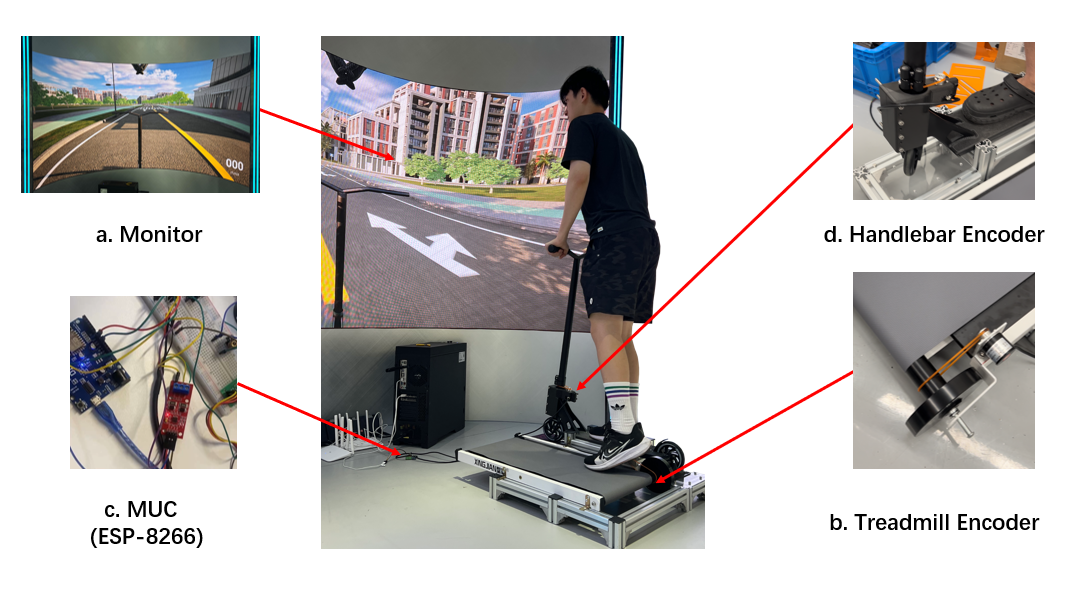}
    \caption{\textsc{NAVIS} system components, which include a modified scooter equipped with a conveyor encoder and a handlebar encoder.}
  \label{fig:NAVIS_teaser}
\end{teaserfigure}



\maketitle

\section{Introduction}

The rapid advancement of virtual reality (VR) technologies has empowered users to immerse themselves in a wide array of digital environments, even when confined to limited physical spaces~\cite{tseng_finger,mueller2021limited}. Despite these strides, navigating expansive virtual worlds remains a significant challenge due to real-world spatial constraints. Traditional VR locomotion techniques, such as teleportation or joystick control, often undermine immersion and realism, leading to diminished spatial awareness and user engagement~\cite{He2024MetaDragonBoat}.

To overcome these limitations, various locomotion methods have been explored. Redirected walking allows users to navigate larger virtual spaces within a confined area by subtly manipulating visual cues, but it can induce motion sickness and requires sophisticated tracking systems \cite{razzaque2005redirected}. Omnidirectional treadmills, like the VirtuSphere developed by Medina et al. \cite{medina2008virtusphere}, enable 360-degree movement and natural locomotion in any direction. While effective in enhancing immersion, such devices are often prohibitively expensive and impractical for widespread use. Alternative approaches utilize common equipment to improve VR locomotion. Sato et al. introduced VibroSkate \cite{sato2015vibroskate}, a system evolved from a skateboard that synchronizes vibration with movement to replicate haptic sensations. Although it offers intuitive control over speed, users find it challenging to manage directional control effectively. Chen et al. developed Slivercycling \cite{chen2024silvercycling}, a natural motion-based locomotion system, to enhance spatial orientation for older people and provide a better user experience. However, Matviienko et al. \cite{Matviienko_2023_cycle} evaluated three VR bicycle simulators and discovered that bicycles are not essential for cycling in VR. Similarly, Deligiannidis et al. developed a VR scooter equipped with haptic features like wind simulation, yet it lacks full physical involvement, which can diminish immersion~\cite{deligiannidis2006vr}.

To address these gaps, we introduce \textsc{NAVIS} (Navigating Virtual Spaces with Immersive Scooters), a novel scooter-based VR system for more natural and integrated navigation \cite{mueller2023toward}. \textsc{NAVIS} captures foot speed via a treadmill and steering through a rotary encoder, closely simulating the experience of riding a scooter outdoors. Integrating haptic feedback and a large CAVE-like screen enhances the immersive experience, creating a realistic sense of movement. \textsc{NAVIS} provides a promising solution for VR locomotion, with potential applications in gaming, education, and more~\cite{gao2023vrprem}.

\section{Implementation}

\begin{figure}[h]
    \centering
    \includegraphics[width=0.5\linewidth]{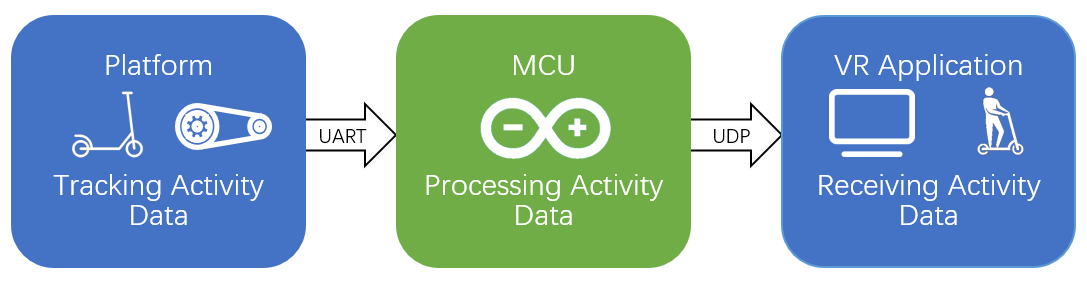}
    \caption{Pipeline for capturing motion data from platform encoders, transmitted every 157 milliseconds via UART to the MCU for processing. The processed data is then synchronized with the VR application via UDP.}
    \label{fig:new1}
\end{figure}

The conceptual framework of \textsc{NAVIS} is composed of three main components, as illustrated in Figure \ref{fig:new1}: (1) a modified indoor scooter-based apparatus, which includes a scooter and a treadmill, (2) an integrated input and transmission system comprising an incremental treadmill encoder, a handlebar encoder, an Arduino-compatible microcontroller unit (MCU), and a WIFI router, and (3) a display system consisting of a PC and a CAVE-like monitor.

\paragraph{\textbf{Scooter-Based Interface}}

To simulate the real-world experience of riding a scooter, we integrated an actual scooter with a treadmill, as illustrated in Figure \ref{fig:NAVIS_teaser}. An incremental rotary encoder was placed near the treadmill to capture user input related to forward and backward movements. Additionally, an absolute encoder was attached to the scooter’s handlebar to gather turning input data, providing precise rotation degree information.

\paragraph{\textbf{Microcontroller Unit}}

The encoder signals were processed by an Arduino-compatible microcontroller (MCU). For forward and backward movement, we applied a clamped linear mapping to treadmill encoder data (in revolutions per second), restricted to a range of -1 to 1. Positive values moved the scooter forward in VR, while negative values moved it backwards. Sliding from the top to the bottom of the treadmill advanced the scooter in VR, with slip speed controlling virtual velocity. For turning, we used the difference in rotational data from the absolute encoder, limited to -180 to 180 degrees. Positive values turned the scooter right, while negative values turned it left in VR by the corresponding degree. This data was transmitted in real-time to a PC via UDP sockets, ensuring synchronized and responsive translation of physical actions to VR. The MCU captured key metrics like treadmill speed and steering angle, minimizing latency.

\paragraph{\textbf{VR Application}}

We utilized the Unreal Engine to create the virtual environment, offering an immersive experience. The VR application is designed to receive the processed data from the MCU and map the physical movements onto the virtual scooter. With the CAVE system displaying the virtual environment that responds in real-time to the kinematic data collected by the platform, \textsc{NAVIS} facilitates indoor physical activities and immersive exploration experiences. Users can drive the virtual scooter anywhere within the virtual environment by moving on the treadmill, leading to a dynamic workout experience that bridges the gap between physical activity and virtual exploration.

\section{Conclusion and Future Work}

The \textsc{NAVIS} system offers a novel approach to navigating virtual environments by integrating real-world movement with immersive digital spaces. Using a scooter-based interface, haptic feedback, and CAVE-like display, \textsc{NAVIS} enhances spatial awareness and intuitive navigation. This approach has applications in gaming, education, and more. Future developments will focus on adding biometric sensors for personalized experiences, particularly in rehabilitation and fitness, and optimizing for urban simulations and obstacle navigation. Planned user studies will evaluate \textsc{NAVIS} in gaming, education, and training contexts. Additionally, scaling for multi-user experiences will enable social interaction and shared exploration in VR.
\begin{acks}
This research is supported by the Center for Aging Science of the Hong Kong University of Science and Technology (Research Program Development in Aging, Healthcare, Mental Wellbeing and COVID Recovery) (Award Number: FS110).
\end{acks}

\bibliographystyle{ACM-Reference-Format}
\bibliography{reference}

\appendix

\end{document}